\documentclass[12pt]{article}

\def\be{\begin{equation}}
\def\ee{\end{equation}}

\usepackage{graphicx,amsmath,amssymb}
\usepackage{multirow}
\usepackage{subfigure}

\begin{document}
\titlepage
\begin{flushright}
IPPP/16/35 \\
\today \\
\end{flushright}

\vspace*{0.5cm}
\begin{center}
{\Large \bf  Sudakov effects in photon--initiated processes}\\

\vspace*{1cm}

L. A. Harland-Lang$^a$, V. A. Khoze$^{b,c}$
and M. G. Ryskin$^c$

\vspace*{0.5cm}
$^a$ Department of Physics and Astronomy, University College London, WC1E 6BT, UK \\           
$^b$ Institute for Particle Physics Phenomenology, Durham University, DH1 3LE, UK    \\
$^c$
 Petersburg Nuclear Physics Institute, NRC Kurchatov Institute, 
Gatchina, St.~Petersburg, 188300, Russia \\
\end{center}

\begin{abstract}
\noindent We consider the effect of the Sudakov factor in photon--initiated processes, corresponding to the no branching probability for the initial--state photon. We demonstrate how such a factor follows simply from the solution of the DGLAP equation for the photon PDF, and is therefore included automatically by this. We use this result to argue that the appropriate scale for the QED coupling $\alpha$ associated with an initial--state photon is not the virtuality of the photon, but rather the factorization scale at which the photon PDF is evaluated, and therefore that the use of the on--shell renormalization scheme is not appropriate for such processes. We also discuss exclusive photon--initiated processes, and demonstrate that no explicit Sudakov factor is required in this case.
\end{abstract}
\vspace*{0.5cm}

\section{Introduction}

As we enter the era of precision LHC phenomenology, where NNLO QCD calculations are becoming the standard for many processes, the influence of electroweak corrections is becoming increasingly relevant. A complete treatment of these inevitably requires the inclusion of diagrams with initial--state photons, with corresponding photon parton distribution function (PDF) introduced in analogy to the more commonly considered PDFs of the quarks and gluons~\cite{Martin:2004dh,Ball:2013hta,Martin:2014nqa,Schmidt:2015zda}. In addition, such photon--initiated processes can lead naturally to exclusive or semi--exclusive final states, where either the colliding protons remain intact after the collision~\cite{Harland-Lang:2015cta}, or there are large rapidity gaps between the dissociating proton systems and the centrally produced state~\cite{Harland-Lang:2016apc}.

While the photon PDF is introduced in exactly the same way as for the other partons within the proton, and obeys a corresponding DGLAP evolution equation, the small size of the QED coupling $\alpha$ leads to some novel results and simplifications which do not occur in the QCD case. In particular, as discussed in~\cite{Harland-Lang:2015cta,Harland-Lang:2016qjy}, the DGLAP equation for the photon PDF can to very good approximation be solved exactly, allowing the solution to be written separately in terms of an input due to low scale coherent and incoherent photon emission from the proton, and an additional term due to high scale emission from the quarks. A crucial element in this separation, which follows naturally from the solution to the evolution equation, is the photon Sudakov factor, corresponding to the probability for the photon to evolve from the starting scale $Q_0$ to the factorization scale $\mu_F$ without further branching. 

In this paper we discuss the effect of this Sudakov factor further, building on the work of~\cite{Harland-Lang:2015cta}, for both inclusive and exclusive production. We will demonstrate that for inclusive processes, where this factor is automatically included via the evolution of the corresponding photon PDF, some care is needed when considering the appropriate renormalization scale at which to evaluate  $\alpha$ for the initial--state photon coupling to the production subprocess. In particular, as the Sudakov factor is generated by photon self--energy diagrams, there is an inevitable overlap with the renormalization of $\alpha$, and we find that the coupling should be evaluated at the factorization scale $\mu_F$ taken for the photon PDF, rather than  the virtuality of the initial--state photon. That is, for processes where a photon PDF has been introduced, it is no longer appropriate to apply the on--shell renormalization scheme, contrary to the approach that is often taken in the literature~\cite{Dittmaier:2009cr,Nhung:2013jta,Baglio:2013toa}, and doing so will lead to an underestimate of the corresponding photon--initiated cross section. 

In exclusive processes, which are given theoretically in terms of the equivalent photon approximation, it may be tempting to introduce such a Sudakov factor by hand, in particular given the crucial role of Sudakov effects in QCD--mediated exclusive processes~\cite{Khoze:1997dr,Khoze:2000cy,Khoze:2001xm}. Here, we clarify the relationship between the equivalent photon approximation and the PDF formalism, and demonstrate that provided the photon virtuality is taken as the scale of the coupling $\alpha$, these effects are automatically included, and there is no need to introduce any explicit Sudakov factor.

The outline of this paper is as follows. In Section~\ref{sud} we demonstrate how the Sudakov factor is generated by the DGLAP evolution of the photon PDF. In Section~\ref{renscale} we consider inclusive production, and argue that the appropriate scale for the coupling $\alpha$ of the initial--state photon to the hard process is the factorization scale of the photon PDF. In Section~\ref{sudex} we consider exclusive processes, and show that there is no need to introduce a photon Sudakov factor by hand, provided the corresponding coupling $\alpha$ is evaluated at the scale of the photon virtuality. In Section~\ref{res} we present some very brief numerical results. Finally in Section~\ref{conc} we conclude.

\section{Evolution of the photon PDF and the Sudakov factor}\label{sud}

The role of the Sudakov factor in photon--initiated processes was recently discussed in~\cite{Harland-Lang:2016apc,Harland-Lang:2016qjy}, and is most easily seen by considering the DGLAP equation for the evolution of the photon PDF $\gamma(x,Q^2)$, which at LO in $\alpha$ and NLO in $\alpha_S$ is given by
\be\label{dglap}
\frac{\partial \gamma(x,Q^2)}{\partial \ln Q^2}=\frac{\alpha(Q^2)}{2\pi}\!\int_x^1\!\frac{dz}z\! \left(P_{\gamma\gamma}(z)\gamma(\frac xz,Q^2)
+\sum_q e^2_qP_{\gamma q}(z)q(\frac xz,Q^2)+ P_{\gamma g}(z)g(\frac xz,Q^2)\right)\;.
\ee
The $P_{\gamma q}(z)$ and $P_{\gamma g}(z)$ are the NLO (in $\alpha_S$) splitting functions, see~\cite{deFlorian:2015ujt}; for simplicity we will work at LO in $\alpha_S$ in this section, before commenting on the NLO case at the end. $P_{\gamma\gamma}$ corresponds to the virtual self--energy correction to the photon propagator, given by
\be\label{pgg}
P_{\gamma\gamma}(z)=-\frac{2}{3}\left[N_c\sum_q e^2_q +\sum_l e^2_l\right]\delta(1-z)\;,
\ee
where $q$ and $l$ denote the active quark and lepton flavours in the fermion loop. This gives a negative contribution to the evolution (\ref{dglap}) which can be interpreted as the decrease in the photon density due to $\gamma \to q\overline{q}$ (or $l^+l^-$) splittings. Indeed, the former process will enter in the evolution of the corresponding quark/anti--quark PDFs, with overall momentum conservation implying
\be\label{mom}
\int {\rm d}z \,z \Big[\sum_{q,\overline{q}} P_{a\gamma}(z) + P_{\gamma\gamma}(z)\Big]=0\;,
\ee
consistent with such an interpretation of the virtual term in (\ref{dglap}).
 
As the virtual correction (\ref{pgg}) is proportional to an overall delta function the corresponding contribution to (\ref{dglap}) is proportional to the photon PDF evaluated at $x$. Therefore, if we ignore the small effect that the photon PDF has on the evolution of the quark and gluons (as discussed in~\cite{Harland-Lang:2016apc}, these generally give less than a 0.1\% correction to the photon), which enter at NLO in $\alpha$, then (\ref{dglap}) can be solved exactly, giving~\cite{Harland-Lang:2016apc}
\be \label{gampdf}
\gamma(x,\mu_F^2)=\gamma(x,Q_0^2)\,S_{\gamma}(Q_0^2,\mu_F^2)+\int_{Q_0^2}^{\mu_F^2}\frac{\alpha(Q^2)}{2\pi}\frac{dQ^2}
{Q^2}\int_x^1\frac{dz}z \;\sum_q e^2_qP_{\gamma q}(z)q(\frac xz,Q^2)\,S_{\gamma}(Q^2,\mu_F^2)\;,
\ee
at LO in $\alpha_S$, where $\gamma(x,Q_0^2)$ is the input PDF at the scale $Q_0$, and we have introduced the photon Sudakov factor, which using (\ref{mom}) can be written as 
\be\label{sudgam}
S_{\gamma}(Q_0^2,\mu_F^2)=\exp\left(-\frac{1}{2}\int_{Q_0^2}^{\mu_F^2}\frac{{\rm d}Q^2}{Q^2}\frac{\alpha(Q^2)}{2\pi}\int_0^1 {\rm d} z\sum_{a}\,P_{a\gamma}(z)\right)\;.
\ee
Here $P_{q(l)\gamma}(z)$ is the $\gamma$ to quark (lepton) splitting function, given by
\be
P_{a\gamma}(z)=N_a\left[z^2+(1-z)^2\right]\;,
\ee
where $N_a=N_c e_q^2$ for quarks and $N_a=e_l^2$ for leptons, while the factor of $1/2$ in (\ref{sudgam}) is present to avoid double counting over the quark/anti--quarks (lepton/anti--leptons). Written in this form, the physical interpretation of the Sudakov factor is clear: it represents the Poissonian probability for no parton emission from the photon during its evolution from the low scale $Q_0$ up to the hard scale $\mu_F$. Thus, the photon PDF (\ref{gampdf}) at $\mu_F$ can be written as the sum of a contribution from low--scale emission of a photon, with no further branching, and a term due to higher scale DGLAP emission from quarks; this separation was used in~\cite{Harland-Lang:2016apc} to demonstrate how a rapidity gap veto can be included in photon--initiated processes. As we will see in Section~\ref{res}, while due to the small size of the QED coupling, $\alpha$, the effect of the Sudakov factor is not dramatic, it is not negligible, in particular for larger evolution lengths.  Finally, we note that the above discussion can be readily generalized to NLO in $\alpha_S$: in this case the splitting function in (\ref{sudgam}) is evaluated up to first order in $\alpha_S$, with a contribution from the $\gamma \to q\overline{q}g$ splitting entering.

\section{The choice of renormalization scale}\label{renscale}

When calculating the cross section for photon--initiated processes we must choose what renormalization scale $\mu_R$ to evaluate the QED coupling $\alpha(\mu_R^2)$ at in the $\gamma\gamma\to X=l^+l^-$, $W^+W^-$... process (or similarly when a single photon is present in the initial--state). Na\"{i}vely, to avoid large higher--order QED corrections we might be tempted to take $\mu_R$ to be of order the hard scale, i.e. $\mu_R\sim \mu_F\sim M_X$, as in the analogous QCD case. However, it is well--known that in QED the appropriate renormalization scale is in general given by the virtuality of the emitted photon: for the $\gamma f\overline{f}$ vertex, the contribution from fermion wave function renormalization and the vertex renormalization exactly cancel due the Ward identity (see e.g.~\cite{Grozin:2005yg} for a pedagogic discussion) so that the only contribution comes from the photon self--energy, with the appropriate scale therefore set by the photon virtuality. Within the collinear factorization approach the initial--state photons are treated as on--shell, and therefore $\alpha$  receives no renormalization at this order, and may be defined in the `on--shell' scheme, i.e. taking $\alpha(\mu_R)=\alpha(0)$. Indeed, such a choice is made frequently in the literature when calculations including photon--initiated contributions are presented, see for example~\cite{Dittmaier:2009cr,Nhung:2013jta,Baglio:2013toa}.

It is our finding that such a choice is in fact inappropriate for processes with initial--state photons, where corresponding photon PDFs have been introduced. To understand why this is the case, we recall that the contribution from the photon self--energy for loop momenta between $Q_0$  and the hard scale $\mu_F$ is already included in the DGLAP evolution of the photon PDF (\ref{dglap}) via the $P_{\gamma\gamma}$ splitting function (\ref{pgg}), and thus care must be taken to avoid double counting when considering the renormalization of $\alpha$. 

To see how this occurs, we can consider for simplicity the contribution of one massless lepton flavour to the photon self--energy. In this case, at one--loop and in the $\overline{{\rm MS}}$ scheme the renormalized scalar part of the self--energy is given by
\be\label{se1}
\Pi(Q^2_0;\mu^2_R)=\frac{\alpha(\mu_R)}{3\pi}\left(\ln\left(\frac{Q^2_0}{\mu^2_R}\right)-\frac{5}{3}\right)
\ee
for renormalization scale $\mu_R$ and a photon of virtuality $-Q_0^2$. However, we can see from (\ref{sudgam}) that this correction for loop momenta between $Q_0$ and $\mu_F$ is already present in the Sudakov factor, and is therefore accounted for by the evolution of the photon PDF. Indeed, we can see that the coefficient of the logarithm is exactly as we would expect from (\ref{sudgam}) after performing the $z$ integration, see (\ref{pgg}).  To avoid double counting we must therefore subtract this from (\ref{se1}), giving
\be\label{se2}
\Pi(Q^2_0;\mu^2_R)\to\frac{\alpha(\mu_R)}{3\pi}\left(\ln\left(\frac{Q^2_0}{\mu^2_R}\right)-\ln\left(\frac{Q^2_0}{\mu_F^2}\right)-\frac{5}{3}\right)=\frac{\alpha(\mu_R)}{3\pi}\left(\ln\left(\frac{\mu_F^2}{\mu^2_R}\right)-\frac{5}{3}\right)\;.
\ee
Thus, while in (\ref{se1}) the natural scale choice is $\mu_R\sim Q_0$, after subtracting the contribution generated by $P_{\gamma\gamma}$, we instead have $\mu_R\sim \mu_F$; if we take $\mu_R\sim Q_0$ this will introduce large $\sim \ln (\mu_F^2/Q_0^2)$ corrections at higher order.

This fact can be seen more clearly if we consider the photon--initiated cross section for some object $X$
\begin{equation}\label{fact}
\sigma(X)=\int {\rm d}x_1{\rm d}x_2 \,\gamma(x_1,\mu^2)\gamma(x_2,\mu^2)\,\alpha(\mu^2)^2\,\tilde{\sigma}(\gamma\gamma\to X)\;.
\end{equation} 
where $\tilde{\sigma}$ is the usual $\gamma\gamma \to X$ subprocess cross section, but with the two powers of $\alpha$ associated with the initial--state photons factored out, and following the discussion above evaluated at a universal scale $\mu_R=\mu_F=\mu$. As discussed above for on--shell external photons we expect the coupling $\alpha$ to receive no renormalization at 1--loop.  If we consider the variation with respect to $\mu$, then for the piece associated with (say) proton 1 we have
\be
\frac{\partial \left(\alpha(\mu^2)\gamma(x_1,\mu^2)\right)}{\partial\ln\mu^2}=\beta_0\,\alpha^2(\mu^2)\gamma(x_1,\mu^2)+\alpha(\mu^2)\,\frac{\alpha(\mu^2)}{2\pi}\left(\int_x^1 \frac{{\rm d}z}{z} P_{\gamma\gamma}(z)\gamma\left(\frac{x}{z},\mu^2 \right)+\cdots\right)
\ee
where we have used the usual expression for QED beta--function at 1--loop and the DGLAP equation (\ref{dglap}), and have left the terms due to photon emission from the quarks implicit. From (\ref{pgg}) we have for the terms associated with the photon self--energy
\be\label{arun}
\frac{\partial \left(\alpha(\mu^2)\gamma(x_1,\mu^2)\right)}{\partial\ln\mu^2}=\alpha^2(\mu^2)\gamma(x_1,\mu^2)\left(\beta_0-\frac{1}{3\pi}\right)=0\;,
\ee
from the known expression for the QED beta--function. Thus the expectation that the cross section has no charge renormalization at 1--loop order due to the photon self--energy is born out, but only if we consistently evaluate the corresponding coupling $\alpha$ at $\mu_R=\mu_F$; if we take the on--shell coupling $\alpha(0)$, while evaluating the photon PDF at a different scale $\mu_F$, the first term in (\ref{arun}) will be absent and this cancellation will no longer occur.

A related argument was in fact made in~\cite{Martin:2006qz}: here, it was shown that if the number of active flavours entering the running of $\alpha_S$ and the evolution of the PDFs is not treated uniformly, then this will lead to discontinuities in physical observables\footnote{We thank Robert Thorne for bringing this to our attention.}, such as the longitudinal structure function, $F_L$. Indeed, such an argument applies here as well. If we consider for example the contribution from the initial--state $\gamma \to q\overline{q}$ splitting to the proton longitudinal structure function
\be
F_{L,\gamma}=\alpha \, C_{L,\gamma}^1 \otimes \gamma\;,
\ee
where $C_{L,\gamma}^1$ is the corresponding coefficient function and `$\otimes$' represents the usual convolution over the momentum fraction $z$, then again we must evaluate $\alpha$ and the photon PDF at the same scale $\mu$, with the same number of active quark (and lepton) flavours in the running of $\alpha$ and the evolution of the photon PDF, through $P_{\gamma\gamma}$, to avoid unphysical discontinuities in $F_L$.

Thus, we are led to conclude that the justification for evaluating the  coupling $\alpha$ at the scale of the photon virtuality does not apply when we consider initial--state photons with corresponding PDFs. While within the collinear approach these incoming photons are treated as on--shell, the use of the on--shell renormalization scheme for $\alpha$, as is often taken in the literature, is inconsistent with the factorization of logarithmic QED corrections in the photon PDF, leading  to double counting of such corrections and to discontinuities in physical observables. The real $\gamma \to q\overline{q}$ splitting will be present in the evolution of the quark/anti--quark PDFs, and from momentum conservation (\ref{mom}) the corresponding virtual quark loop contributions to the photon propagator must therefore also be included in the photon evolution.  Physically,  the photon substructure is in effect being resolved, such that the use of a purely on--shell scheme is no longer appropriate; the photon self--energy contribution can never be consistently fully absorbed into the coupling $\alpha$, as this must be explicitly present in the photon PDF evolution.

\section{The Sudakov factor in exclusive processes}\label{sudex}

In the Section~\ref{sud} we introduced the Sudakov factor (\ref{sudgam}) within the standard collinear factorization formalism that is used to calculate the inclusive cross section for the $\gamma\gamma$--initiated production of a system $X$ accompanied by an arbitrary number of additional particles. However, we may also consider exclusive or semi--exclusive processes, which are naturally generated in $\gamma\gamma$--initiated production: in the former case where the protons remain intact after the collision, and in the latter where a veto is imposed on additional particle production in a large rapidity interval. Indeed, semi--exclusive processes have been considered in~\cite{Harland-Lang:2016apc}, with the separation achieved in (\ref{gampdf}) through the introduction of the Sudakov factor being crucial in the derivation of an `effective' photon PDF, modified by the rapidity veto.

It is therefore also natural to consider the potential impact of the Sudakov factor on purely exclusive processes, that is the production of an object $X$ via
\be
pp\to p\,+\,X\,+\,p\;,
\ee
where the protons remain intact after the collision, and the `+' correspond to large rapidity gaps between the outgoing intact protons and object $X$. Indeed in QCD--initiated processes a Sudakov factor, corresponding to the probability for no emission from the fusing gluons, is known to play crucial role, ensuring that the calculation itself is perturbatively stable~\cite{Khoze:1997dr,Khoze:2000cy,Khoze:2001xm}. While there is no question of perturbative stability for the purely QED processes we consider here, we may nonetheless expect the corresponding photon Sudakov factor to contribute.

For purely exclusive processes it is no longer necessary to introduce a photon PDF, obeying a corresponding DGLAP evolution equation. Rather, the exclusive cross section is given in the equivalent photon approximation~\cite{Budnev:1974de} in terms of the number density of quasi--real photons emitted by the proton
\begin{equation}\label{gamcoh}
n(x_i)=\frac{1}{x_i}\frac{\alpha}{\pi}\int_0^{Q^2_i<Q_0^2}\!\!\frac{{\rm d}q_{i_t}^2 }{q_{i_t}^2+x^2_i m_p^2}\left(\frac{q_{i_t}^2}{q_{i_t}^2+x^2_i m_p^2}(1-x_i)F_E(Q^2_i)+\frac{x^2_i}{2}F_M(Q^2_i)\right)\;,
\end{equation}
where  $F_E$ and $F_M$ are given in terms of the usual proton electric and magnetic form factors, and $Q_0$ is some upper limit on the photon virtuality, see~\cite{Harland-Lang:2016apc} for further details. On the other hand, in the PDF formalism the photon density at the starting scale $Q_0$ is written as~\cite{Harland-Lang:2016apc} (see also~\cite{Martin:2014nqa})
\be
\gamma(x,Q_0^2)=\gamma^{\rm coh}(x,Q_0^2)+\gamma^{\rm incoh}(x,Q_0^2)\;,
\ee
that is as a sum of terms due to coherent and incoherent low--scale photon emission from the proton; comparing this with the EPA framework we have $n(x_i)=\gamma^{\rm coh}(x_i,Q_0^2)$. From the discussion in Section~\ref{sud} we know that it is this PDF at the starting scale multiplied by the corresponding Sudakov factor that is the relevant object in the high scale $\gamma\gamma \to X$ cross section. Indeed, in~\cite{Harland-Lang:2016apc}, the contribution from this input component is expected to automatically pass a rapidity veto in the semi--exclusive case, but such a conclusion is only physically justified after the inclusion of this factor, which allows the separation in (\ref{gampdf})  to be achieved. Thus, for purely exclusive processes it is tempting  to conclude that such a Sudakov factor should be included, i.e. that we should multiply the equivalent photon density in (\ref{gamcoh}) by this.

However, following the discussion in Section~\ref{renscale} we find that this is not the case. In particular, we found there that  for inclusive processes which automatically include such a Sudakov factor, through the DGLAP evolution of the corresponding photon PDFs, the appropriate scale for the corresponding coupling $\alpha$ associated with the hard process is not the photon virtuality but rather the factorization scale $\mu_F$ of the PDF. In this case, we find that as expected the cross section contains no scale  dependence at 1--loop due to the photon self--energy correction, see (\ref{arun}). On the other hand for exclusive processes, the standard EPA formulae does not include a Sudakov factor, and so this argument no longer applies: rather, we may absorb the entirety of the photon self--energy correction into the renormalization of $\alpha$, and the natural scale to take is the photon virtuality. Moreover, if  in (\ref{sudgam}) we take $\alpha(Q^2)\approx \alpha(Q_0^2)$, which is valid up to higher--order corrections, then considering the contribution from one lepton flavour for simplicity, we have
\begin{align}\nonumber
S_{\gamma}(Q_0^2,\mu_F^2)&=\exp\left(-\frac{1}{2}\int_{Q_0^2}^{\mu_F^2}\frac{{\rm d}Q^2}{Q^2}\frac{\alpha(Q^2)}{2\pi}\int_0^1 {\rm d} z\,P_{l\gamma}(z)\right)\;,\\
&\approx \exp\left(-\alpha(Q_0^2)\,\beta_0\ln\left(\frac{\mu_F^2}{Q_0^2}\right)\right)\;,
\end{align}
where $\beta_0=1/3\pi$ is the leading coefficient of the QED beta--function. This implies that
\be\label{sudalpha}
\alpha(\mu_F^2)S_{\gamma}(Q_0^2,\mu_F^2)=\alpha(\mu_F^2)\exp\left(\frac{\alpha(Q_0^2)}{\alpha(\mu_F^2)}-1\right)\approx \alpha(Q_0^2)\;,
\ee
up to higher order terms in $\alpha$. Thus at 1--loop level the inclusion of a Sudakov factor and evaluation of the coupling at scale $\mu_F$ is  exactly equivalent to simply taking the photon virtuality as the scale of the coupling, with the Sudakov factor omitted\footnote{In (\ref{sudalpha}) the scale $Q_0$ in fact corresponds to the upper limit on the flux (\ref{gamcoh}) integrated over the photon virtuality, however an identical argument can be made for the unintegrated flux, for which the low scale $Q_0$ corresponds to the photon virtuality itself.}. Therefore, in exclusive  $\gamma\gamma$--initiated processes, there is no need to introduce a Sudakov factor; due to the particular form of the QED charge renormalization, the `no--emission' probability that plays such a crucial role in the QCD--mediated case is automatically accounted by simply evaluating the coupling $\alpha$ at the scale of the photon virtuality. Thus, when comparing the predictions for exclusive lepton pair production as in~\cite{Harland-Lang:2015cta} to data from ATLAS~\cite{Aad:2015bwa} we can expect no further reduction in the predicted cross section coming from the inclusion of a Sudakov factor.

It is worth commenting again on the comparison to the QCD case, for which the Sudakov factor, in contrast, plays such a crucial role~\cite{Khoze:1997dr,Khoze:2000cy,Khoze:2001xm}. The fundamental difference is that in QED, for which there is no 3-photon vertex, the corresponding Sudakov factor is only a single--logarithmic function. That is, it is given in terms of the multiplicity of $q\overline{q}$ pairs emitted during the photon DGLAP evolution, which is only enhanced by a single logarithm; in terms of Feynman diagrams this is given by quark self--energy insertions to the photon propagator. Thus in QED the Sudakov factor may be compensated by an appropriate choice of scale for $\alpha$, for which the renormalization is driven by precisely the same quark loop insertions. On the other hand, in QCD the presence of the $g\to gg$ transition generates a double logarithmically enhanced gluon multiplicity which can no longer be compensated in this way.

\section{Results}\label{res}

 \begin{figure}[h]
\begin{center}
\includegraphics[scale=0.65]{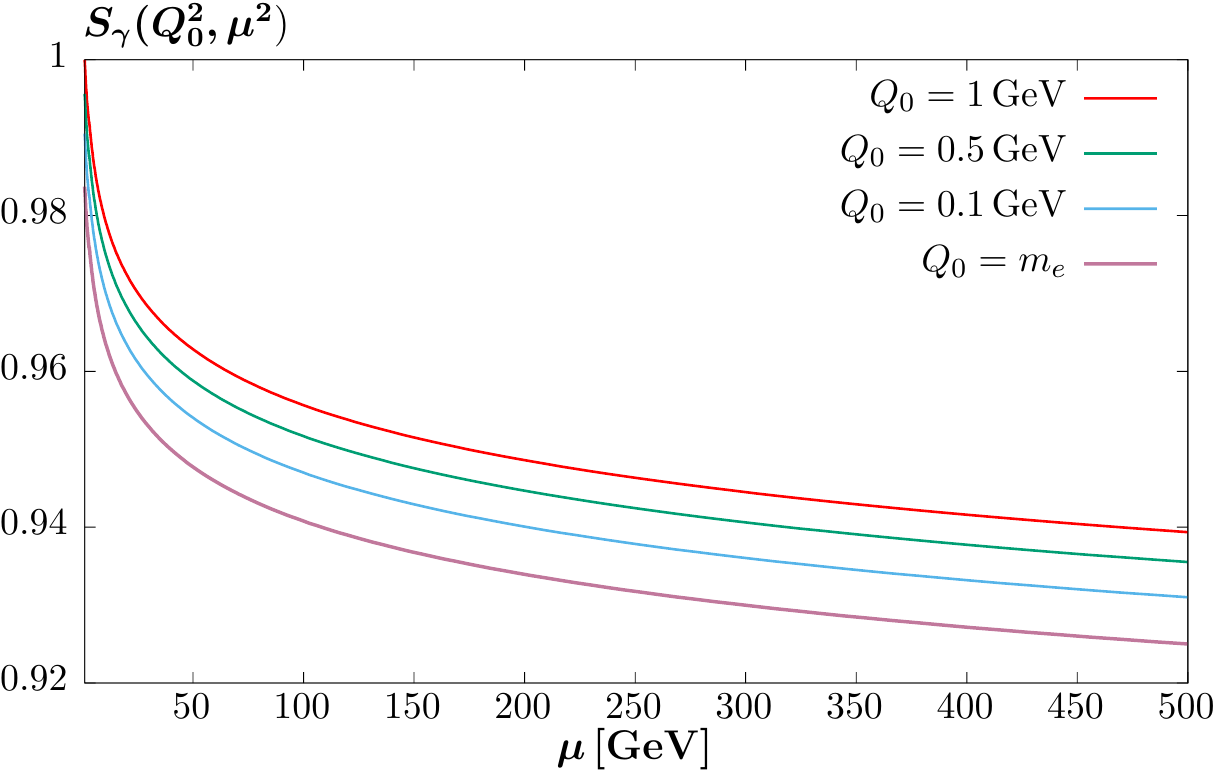}
\caption{Photon Sudakov factor $S_\gamma(Q_0^2,\mu^2)$ as a function of the hard scale $\mu$ for different values of the modulus of the photon virtuality $Q_0^2$.}
\label{fig:sud1}
\end{center}
\end{figure} 

 \begin{figure}[h]
\begin{center}
\includegraphics[scale=0.65]{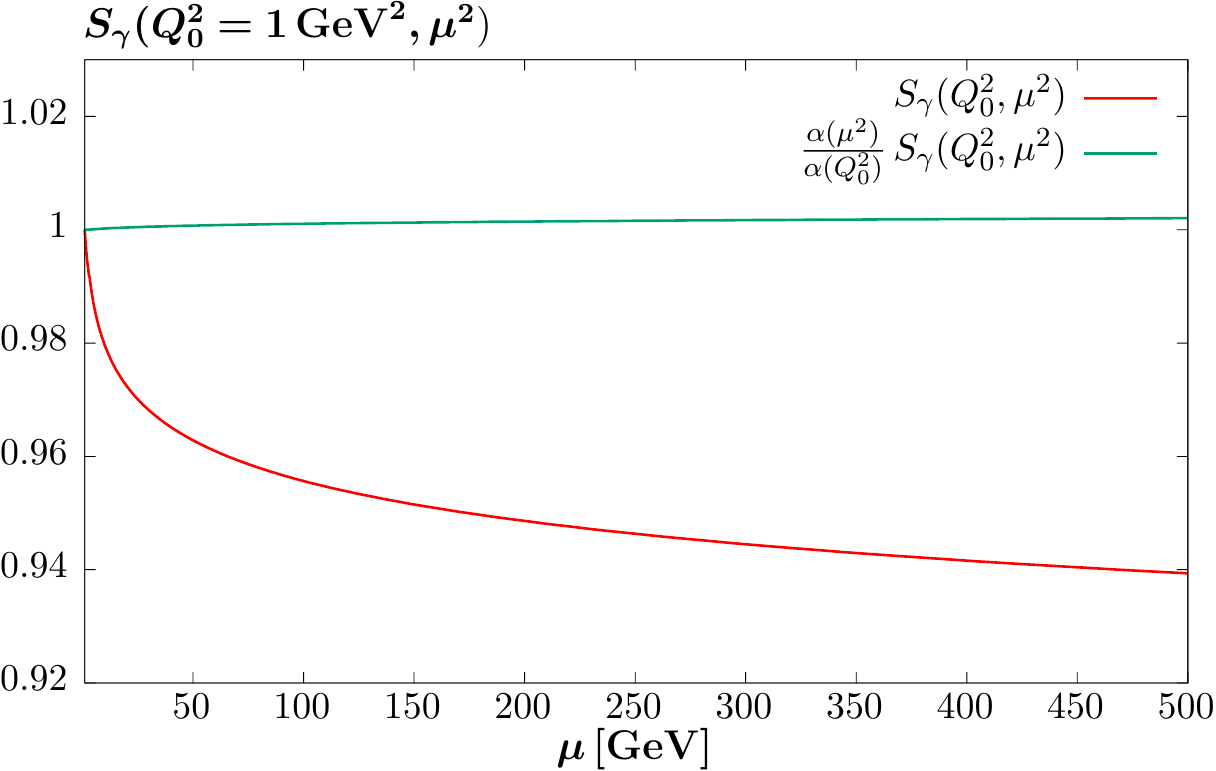}
\caption{Photon Sudakov factor $S_\gamma(Q_0^2,\mu^2)$ for $Q_0^2=1\,{\rm GeV}^2$ as a function of the hard scale $\mu$, and multiplied by the ratio $\alpha(\mu^2)/\alpha(Q_0^2)$.}
\label{fig:sud2}
\end{center}
\end{figure} 

In Fig.~\ref{fig:sud1} we show the photon Sudakov factor (\ref{sudgam}) as a function of the upper scale $\mu$, for a range of choices of lower scales $Q_0$. We can see that the suppression is, as expected, quite small but not negligible, with a reasonably gentle dependence on  $Q_0$. In Fig.~\ref{fig:sud2} we show the Sudakov factor for $Q_0=1$ GeV, again as a function of $\mu$, but in addition show the case where this is multiplied by $\alpha(\mu^2)$, with an additional factor of $\alpha(Q_0^2)$ included in the denominator so that the result is normalized to unity at $\mu=Q_0$. We can see, as expected from (\ref{sudalpha}) that the latter quantity is extremely close to unity, up to the $\lesssim 0.1\%$ level (consistent with residual $O(\alpha^2)$ and higher corrections) and is almost constant across a large range of $\mu$. This supports the conclusion of Section~\ref{sudex}, namely that we do not need to include such a Sudakov factor in exclusive processes. On the other hand, for inclusive processes, where this is automatically included via the evolution of the photon PDF, from Fig.~\ref{fig:sud2}  we can see that there will be an unphysical $\sim S_\gamma^2\sim 0.9$ suppression in the predicted $\gamma\gamma$--initiated cross section if the photon virtuality is taken as the scale of $\alpha$, i.e. if the on--shell scheme is used. On the other hand, if $\alpha$ is evaluated consistently at the factorization scale $\mu_F$ we can see that the renormalization scale dependence of the cross section is negligible, as expected from (\ref{arun}).

\section{Conclusion}\label{conc}

Photon--initiated processes play an important role in both inclusive and exclusive production at the LHC. In the former case, these must inevitably be introduced as part of electroweak corrections, which are becoming increasingly topical in light of the high level of precision required for much LHC phenomenology. In the latter, initial--state photons can naturally lead to exclusive or semi--exclusive final states, where the colliding protons remain intact after the collision, or there are large rapidity gaps between the proton dissociation products and the centrally produced system. 

To calculate the cross section for inclusive photon--initiated processes, it is necessary to introduce a photon PDF, in exact analogy to the other partons within the proton. This obeys an equivalent DGLAP evolution equation but which, in contrast to the quarks and gluons, is generated to first order by purely QED emission and virtual correction diagrams. In such a case the small size of the QED coupling $\alpha$ allows a simple solution  for the photon PDF to be written down  to very good accuracy, given separately in terms of an input due to low scale coherent and incoherent photon emission from the proton, and high scale emission from the quarks. Such a separation allows us to treat exclusive, semi--exclusive and inclusive photon--initiated processes within the same framework, with different components of the same photon PDF being probed in different cases. Crucial in achieving this is the introduction of a Sudakov factor, corresponding to the probability of no photon branching between the starting scale $Q_0$ and the high scale $\mu_F$ of the production subprocess.

In this paper we have examined in detail the role that this Sudakov factor plays in both inclusive and exclusive photon--initiated processes. We have shown that, as this is automatically generated by the the photon self--energy contribution to the DGLAP evolution of the photon PDF, there is an inevitable overlap with the renormalization of the corresponding QED coupling $\alpha$ of the initial--state photon to the production subprocess. This implies that  the factorization scale $\mu_F$, and not the photon virtuality, should be chosen as the scale of the coupling, in contrast to the approach that is most often taken in the literature. On the other hand, in the exclusive case we have verified that that no explicit Sudakov factor needs to be introduced, as the effect of this is compensated by evaluating $\alpha$ at the scale of the photon virtuality. These results therefore guide the approach that should be taken when considering both inclusive and exclusive photon--initiated processes at the LHC, in particular when high precision is desired.

\section*{Acknowledgements}

We thank Robert Thorne for useful discussions. The work of MGR  was supported by the RSCF grant 14-22-00281. VAK thanks the Leverhulme Trust for an Emeritus Fellowship.  LHL thanks the Science and Technology Facilities Council (STFC) for support via the grant award ST/L000377/1. MGR thanks the IPPP at the University of Durham for hospitality.

\bibliography{references}{}
\bibliographystyle{h-physrev}

\end{document}